\documentclass[12pt]{amsart}
\usepackage{graphicx}

\newtheorem{thm}{Theorem}
\newtheorem{lem}{Lemma}[section]

\newtheorem{prop}[lem]{Proposition}
\theoremstyle{definition}

\theoremstyle{remark}
\newtheorem{rem}{Remark}[section]
\numberwithin{equation}{section}

\newcommand{\set}[1]{\left\{#1\right\}}

\newcommand{\calU}{\mathcal{U}}

\newcommand{\bbZ}{\mathbb{Z}}
\newcommand{\Z}{\mathbb{Z}}
\newcommand{\Q}{\mathbb{Q}}
\newcommand{\R}{\mathbb R}

\newcommand{\bbT}{\mathbb{T}}

\newcommand{\Op}{ \mbox{Op}}
\newcommand{\SL}{ \mbox{SL}}

\newcommand{\Sp}{\mathrm{Sp}}

\newcommand{\calH}{\mathcal{H}}

\begin{document}
\title[Quantum variance on the 4-dimensional torus ]
{On the Quantum variance of matrix elements for the cat map on the 4-dimensional torus}%
\author{DUBI KELMER}%
\address{Raymond and Beverly Sackler School of Mathematical Sciences,
Tel Aviv University, Tel Aviv 69978, Israel}
\email{kelmerdu@post.tau.ac.il}

\thanks{}%

\date{}%
\dedicatory{}%
\commby{}%

\begin{abstract}
    For many classically chaotic systems, it is believed that in
    the semiclassical limit, the matrix elements of smooth
    observables approach the phase space average of the
    observable. In the approach to the limit the matrix elements
    can fluctuate around this average.
    Here we study the variance of these fluctuations, for the
    quantum cat map on $\mathbb{T}^4$.
    We show that for certain maps and observables, the variance has a different
    rate of decay, than is expected for generic chaotic
    systems.
\end{abstract}

\maketitle

\section{Introduction}
    The basic question of ``Quantum Chaos''
    is what signatures of the classical (chaotic) dynamics
    one can still find in the quantized system.
    One important such signature is
    given by the asymptotic behavior of expectation values, and
    the limit distribution around their average in the semiclassical limit.

    For classically chaotic systems, the average of smooth observables on phase
    space, along the trajectories of the system, tend almost always to the phase space average
    (i.e., the classical dynamics is ergodic).
    The quantum analog of this, following the correspondence principle,
    is that the expectation values of an observable
    should tend to the phase space average of the
    observable in the semiclassical limit.
    The first result in this direction is ``\v{S}hnirel'man's theorem'', which states that
    at least in the sense of convergence in the mean square, when taking Planck's constant $h\rightarrow 0$,
    the matrix elements converge to the phase space average ~\cite{BD,DD,S,Z}.
    This notion is usually referred to as ``Quantum Ergodicity" (Q.E.).
    \footnote{The case where this convergence holds for all eigenstates is
    referred to as ``Quantum Unique Ergodicity" (Q.U.E.).}

    In the approach to the limit,
    the different matrix elements can fluctuate about their
    classical limit. In ~\cite{FP} Feingold and Peres proposed a
    formula for the variance of the fluctuations in the semiclassical limit.
    As mentioned above, according to \v{S}hnirel'man's theorem the variance vanishes as
    $h\rightarrow 0$. Nevertheless, after normalizing by an appropriate
    power of $h$, the Feingold-Peres formula relates the variance of
    the matrix elements to classical correlations of the
    observable.
    In ~\cite{EFK1}, Eckhardt et al. developed a semiclassical theory for the variance of these
    fluctuations, giving support for the validity of the
    Feingold-Peres formula, and suggesting that after normalizing by the
    variance,
    the fluctuation should be Gaussian (at least for hyperbolic systems).
    The analysis in \cite{EFK1}, for the fluctuations of the matrix elements,
    can be extended to deal with quantum maps ~\cite{CKR1}.
    In particular, for generic hyperbolic quantum maps on the 2d-torus $\mathbb{T}^{2d}$,
    it predicts Gaussian distribution with variance of order
    $\frac{\alpha }{N^d}$,
    where $N=1/h$ plays the role of the inverse of Planck's constant, and $\alpha$
    is related to the classical variance of the observable.

    A fundamental example for a quantum map on the torus is the quantum cat map,
    originally introduced by Hannay and Berry \cite{HB}.
    Here the classical dynamics is simply the iteration of a
    symplectic linear map $A\in \Sp(2d,\bbZ)$ on $\bbT^{2d}=\R^{2d}/\bbZ^{2d}$.
    When $A$ has no roots of unity for eigenvalues, the
    classical dynamics is `chaotic' (i.e., ergodic and mixing).

    For the quantization of the cat map, the admissible values of
    Planck's constant are inverses of integers, and the space of
    quantum states is then $\calH_N=L^2[(\bbZ/N\bbZ)^d]$.
    The semiclassical limit is achieved by taking $N\rightarrow
    \infty$, and we restrict the discussion to the case where $N$ is prime.
    For $f\in C^\infty(\bbT^{2d})$  a smooth observable, we denote by
    $\Op_N(f):\calH_N\rightarrow \calH_N$ its quantization.
    The quantization of the map $A$, is then a unitary operator $U_N(A)$
    acting on $\calH_N$.

    For $d=1$, this model was studied extensively and it exhibits some interesting features
    \cite{HB,FND,KR1,KR2}.
    It is shown to be (Q.E.), but not (Q.U.E.).
    \footnote{Faure, Nonnenmacher and ~De Bi\`{e}vre constructed a sparse subsequence of eigenstates where the
    expectation values do not converge to the phase space
    average \cite{FND}.} However, in \cite{KR2} Kurlberg and Rudnick introduced
    a group of symmetries of the system,
    a family of commuting unitary maps of $\calH_N$ that commute with
    $U_N(A)$.
    These operators are called Hecke operators, in analogy to a similar setup on the modular surface \cite{Iw}.
    The space $\calH_N$ has an orthonormal basis consisting of
    joint eigenstates $\{\phi_j\}_{j=1}^N$ called ``Hecke
    eigenstates''. For these states the system is shown to be (Q.U.E.)~\cite{KR2}.

    In \cite{KR1} Kurlberg and Rudnick studied the fluctuation of
    the matrix elements for the desymmetrized system
    and gave a conjecture for the limit distribution,
    which is radically different from the behavior expected in the generic case.
    In particular, a fourth moment calculation for the distribution
    showed that it is not Gaussian.
    Furthermore, though the variance of the fluctuation is of order $\frac{1}{N}$
    as expected, the classical factor is different from
    the classical variance, in contrast to the Feingold-Peres formula
    (for comparison see \cite{KR1}).
    \begin{rem}
    In \cite{LS2} Luo and Sarnak showed similar behavior
    of the variance for Hecke eigenforms on the modular surface.
    For this system, the variance of the fluctuations after
    appropriately normalizing, does not converge to the classical
    variance of the observable. However the classical variance can be
    recovered, after inserting some arithmetic correction.
    Furthermore the limit distribution for the Hecke eigenforms is again not
    Gaussian (pending on conjectures of Keating and Snaith \cite{KeS}).
    \end{rem}

    In this note we look at the matrix elements for quantized cat maps on
    $\mathbb{T}^{4}$, and the variance of their fluctuations.
    Here we find that for certain maps and observables,
    the variance has a different rate of decay than that predicted by the
    Feingold-Peres formula.

    Let $A\in \Sp(4,\Z)$ be a matrix with $4$ distinct eigenvalues. For the dynamics to be
    ergodic we assume that $A$ has no roots of unity for
    eigenvalues. We further assume that the vector space $\Q^4$
    decomposes into two rational orthogonal symplectic subspaces,
    invariant under the action of $A$. Denote by $Z_1,Z_2$, the lattices obtained
    by intersecting these subspaces with $\bbZ^4$.

    Analogously to the treatment of the case $d=1$ we introduce Hecke operators, a
    family of commuting operators that commute with $U_N(A)$, and consider a basis of joint
    eigenstates $\set{\phi_j}_{j=1}^{N^2}$, referred to as the Hecke basis.
    For any smooth observable $f\in C^\infty(\bbT^4)$, denote the quantum
    variance in the Hecke basis by
    \[S_2(f,N)=\frac{1}{N^2}\sum_{j=1}^{N^2}\bigg|\langle \Op_N(f)\phi_j,\phi_j\rangle-\int_{\bbT^4}f\bigg|^2.\]

    \begin{thm}\label{t:1}
    Let $f(\vec{x})=e^{2\pi i \vec{n}\cdot\vec{x}},\;\vec{n}\neq
    0$.
    Then, as $N\rightarrow \infty$ through primes, the variance of
    these observables is given by
    \[S_2(f,N)=\left\lbrace\begin{array}{cc}
    \frac{1}{N}+O(\frac{1}{N^2}) & \mbox{if } \vec{n}\in Z_1\cup Z_2,\\
     \frac{1}{N^2}+O(\frac{1}{N^3}) & \mbox{ otherwise }.\
    \end{array}\right.\]
    \end{thm}
    This result is in contrast to the expected rate of
    $\frac{1}{N^2}$, predicted to hold for all observables in generic hyperbolic systems.

    \begin{rem}
    This result can be extended to any smooth
    observable $f\in C^{\infty}(\bbT^{4})$, to state that
    $S_2(f,N)=\frac{V_1(f)}{N^2}+O(\frac{1}{N^3})$, if the Fourier coefficients of $f$
    are supported outside the lattices $Z_1\cup Z_2$, and
    $S_2(f,N)=\frac{V_2(f)}{N}+O(\frac{1}{N^2})$ otherwise.
    The terms $V_1(f),V_2(f)$ can be expressed as quadratic forms in the Fourier
    coefficients of $f$, see \cite{K} for more details.
    \end{rem}

\begin{figure}[h]
            \includegraphics[width=3.2in ]{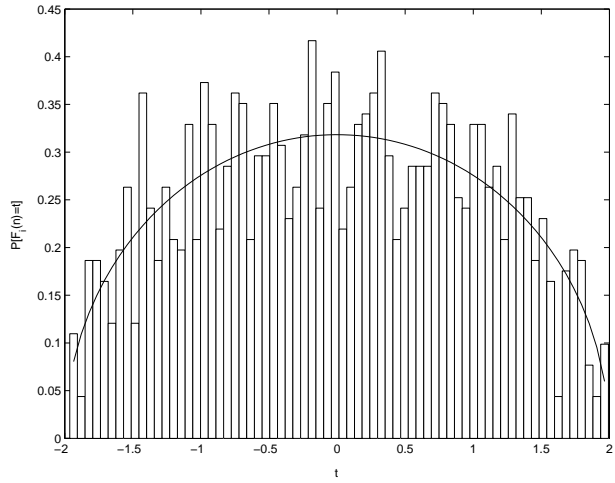}
            \caption{  Scaled distribution of normalized matrix elements
            $ F_i(\vec{n})=\sqrt{N}\langle T_N(\vec{n})\phi_i,\phi_i\rangle\;,\;\vec{n}=(1,0,0,0)\in Z_1$ for $N=1613$.
            Plotted together with a semi circle distribution $y(t)=\frac{1}{2 \pi}\sqrt{4-t^2}$.}\label{f:1}
    \end{figure}

\begin{figure}[h]
            \includegraphics[width=3.2in ]{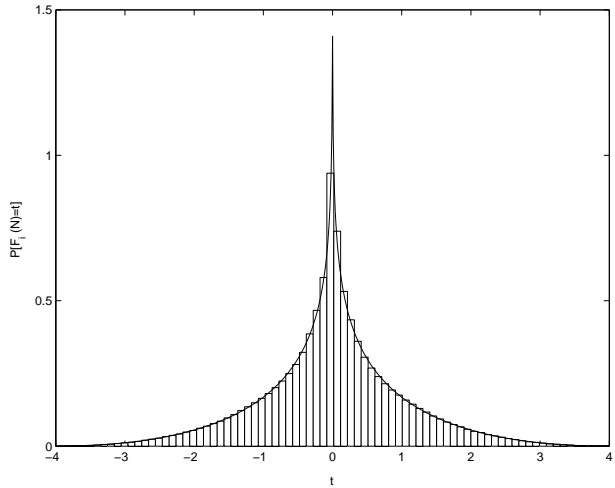}
            \caption{  Scaled distribution of normalized matrix elements
            $F_i(\vec{n})=N\langle T_N(\vec{n})\phi_i,\phi_i\rangle\;,\;\vec{n}=(1,1,0,0)\not\in Z_1\cup Z_2$ for $N=1613$.
            Plotted together the corresponding distribution $y(t)=\frac{1}{2
            \pi^2}\int_{t/2}^2\sqrt{(4-s^2)(4-(\frac{s}{t})^2)}\frac{ds}{s}$}\label{f:2}
    \end{figure}

    After knowing the variance of the fluctuation, we can
    renormalize the matrix elements and consider the limit distribution.
    We present some numerical calculations of this distribution,
    for the two types of elementary observables (figures \ref{f:1},\ref{f:2} respectively).
    The numerical evidence suggests that when $\vec{n}\in Z_1\cup Z_2$, the
    distribution of the normalized matrix elements is given by the semicircle law,
    where otherwise after appropriately normalizing, it corresponds to
    the product of two semicircle random variables. This result, is in
    agreement with the conjecture presented in \cite{KR1} for the
    limit distribution for the cat map on $\bbT^2$.

    Furthermore, assuming that all matrix elements are roughly of the same order,
    the square root of the variance obtained here, gives an
    estimate of the rate of their decay. In figure \ref{f:3}, we
    present numerical calculation for this decay,
    strongly supporting the validity of this estimate.

 \section*{acknowledgments}
  I would like to thank my Ph.D. advisor Zeev Rudnick for his advice and guidance.
  I would also like to thank Par Kurlberg for long discussions of his work.

  This work was supported in part by the EC TMR network
  \textit{Mathematical Aspects of Quantum Chaos}, EC-contract no
  HPRN-CT-2000-00103 and the Israel Science Foundation founded by
  the Israel Academy of Sciences and Humanities. This work was
  carried out as part of the author's Ph.D thesis at Tel Aviv
  University, under the supervision of prof. Zeev Rudnick. A
  substantial part of this work was done during the author's visit
  to the university of Bristol.

\section{Background}
    The full details for the cat map and it's quantization can
    be found in ~\cite{KR2} for one dimensional system, and in \cite{BD1,K} for higher dimensions.
    We briefly review the setup for the quantization of the cat map on $\bbT^4$.
    \subsection{Classical dynamics}
    The classical dynamics are given by the iteration of a
    symplectic linear map $A\in \Sp(4,\bbZ)$.
    \[\vec{x}=\left(%
    \begin{array}{c}
    \vec{p} \\
    \vec{q} \\
    \end{array}%
    \right)\in\mathbb{T}^4\mapsto A\vec{x}.\]
    Given an observable $f\in
    C^\infty(\mathbb{T}^4)$, the classical evolution defined by $A$ is
    $f\mapsto f\circ A$. If $A$ has no eigenvalues that are roots
    of unity then the classical dynamics is ergodic and mixing \cite{Knauf}.
    \subsection{Quantum kinematics}
    For doing quantum mechanics on the torus, one takes Planck's
    constant to be $1/N$, as the Hilbert space of states one takes
    $\mathcal{H}_N=L^2((\bbZ/N\bbZ)^2)$, where the inner product is given
    by:
    \[\langle\phi,\psi\rangle=\frac{1}{N^2}\sum_{\vec{y} \mod{N}}\phi(\vec{y})\overline{\psi(\vec{y})}.\]
    For $\vec{n}=(\vec{n}_1,\vec{n}_2)\in \bbZ^4$ (where $\vec{n}_i\in \bbZ^2$), define elementary operators $T_N(\vec{n})$ acting
    on $\psi\in \mathcal{H}_N$ via:
    \begin{equation}\label{eTN:1}
    T_N(\vec{n})\psi(\vec{y})=e_N(\frac{\vec{n}_1\cdot\vec{n}_2}{2})e_N(\vec{n}_2\cdot\vec{y})\psi(\vec{y}+\vec{n}_1^t),
    \end{equation}
    where $e_N(x)=e^{\frac{2\pi i x}{N}}$

    For any smooth classical observable $f\in C^\infty(\mathbb{T}^4)$
    with Fourier expansion
    $f(\vec{x})=\sum_{\vec{n}\in\bbZ^4}\hat{f}(\vec{n})e^{2 \pi i\vec{n}\cdot\vec{x}}$,
    its quantization is given by
    \[Op_N(f)=\sum_{\vec{n}\in\bbZ^4}\hat{f}(\vec{n})T_N(\vec{n}).\]

    The main properties of the elementary operators $T_N(\vec{n})$
   are summarized in the following proposition.
   \begin{prop}\label{pTN:1}
     For the operators $T_N(\vec{n})$ defined above:
     \begin{enumerate}
     \item $T_N(\vec{n})$ only depends on $\vec{n}\mod{2N}$.
     \item $T_N(\vec{n})$ are unitary operators.
     \item The composition of two elementary operators is
       \[T_N(\vec{m}) T_N(\vec{n})=e_{2N}(\omega(\vec{m},\vec{n})) T_N(\vec{n}+\vec{m}),\]
       where $\omega(\vec{m},\vec{n})=\vec{m}_1\cdot\vec{n}_2-\vec{m}_2\cdot\vec{n}_1$, is the symplectic
       inner product.
     \item If $\vec{n}=0\pmod{N}$, then
     $T_N(\vec{n})=(-1)^{(\vec{n}_1\cdot{\vec{n}_2})}I$.
     \end{enumerate}
   \end{prop}
   These properties are easily derived from equation \ref{eTN:1}.
   Moreover, these properties uniquely characterize
   these operators, in the sense that if $\widetilde{T}_N(\vec{n})$
   are operators acting on an $N^2$-dimensional Hilbert space
   fulfilling the above properties, then they are unitarily
   equivalent to $T_N(\vec{n})$.

    \subsection{Quantum dynamics:} For $A\in \Sp(4,\bbZ)$ which satisfies
    certain parity conditions (namely $A=I\!\pmod{2}$), we assign
    unitary operators $U_N(A)$, acting on $L^2((\bbZ/N\bbZ)^2)$ having the
    following important properties:
    \begin{enumerate}
    \item ``Exact Egorov": For all observables $f\in
    C^\infty(\mathrm{\bbT}^4)$
    \[U_N(A)^{-1}Op_N(f)U_N(A)=Op_N(f\circ A).\]
    \item The quantization depends on $A$ modulo $2N$.
    \item The quantization preserves commutation relations:\\ if $AB=BA\!\pmod{2N}$
    then $U_N(A)U_N(B)=U_N(B)U_N(A)$.
    \end{enumerate}
     Note that if we multiply the operators by arbitrary phases, then
    all the above properties still hold. In fact the converse
    also holds, that is if $\tilde{U}_N(A)$ is an operator
    satisfying these properties, then $\tilde{U}_N(A)=e^{i\alpha} U_N(A)$.

    \begin{rem}
    Eventually, we are interested in the eigenfunctions
    of these operators, which are not affected by the choice of
    phase. Thus, in order to simplify the discussion we do not set
    explicit phases for the quantization.
    However, we note that one can make a specific choice of phases for which this
    quantization is multiplicative, $U_N(AB)=U_N(A)U_N(B)$. In particular this
    implies property (3) above, indeed holds for any choice of phases.
    \end{rem}

    Since we assume that $N$ is odd, and the quantization factors
    through the group
    \[\set{B \in \Sp(4,\bbZ/2N\bbZ)|B=I\!\!\!\!\!\pmod{2}}\simeq \Sp(4,\bbZ/N\bbZ),\]
    we can define a (projective) representation of
    $\Sp(4,\bbZ/N\bbZ)$:\\
    For any element $B\in\Sp(4,\bbZ/N\bbZ)$, let $\bar{B}\in \Sp(4,\bbZ/2N\bbZ)$ be the unique
    element s.t. $B=\bar{B} \pmod{N},\;\;\bar{B}=I\pmod{2}$, and identify $U_N(B):=U_N(\bar{B})$.
    For the rest of this note we will use this convention for the
    quantization of $\Sp(4,\bbZ/N\bbZ)$.

    \subsection{Hecke eigenfunctions}
        Let $A\in \Sp(4,\bbZ)$, $A=I\!\!\pmod{2}$
        and let $C(N,A)$ be a maximal commutative subgroup of
        $\Sp(4,\bbZ/N\bbZ)$, which includes $A$ modulo $N$.
        \begin{rem}
        In general, this
        group is not necessarily unique. However if all the
        eigenvalues of $A\!\!\!\mod{N}$ are distinct then $C(N,A)$ is
        unique. For an explicit construction see \cite{K}.
        \end{rem}
        Since the quantization preserves
        commutation relations, the operators $U_N(B)$, $B\in C(N,A)$
        form a family of commuting operators, called Hecke
        operators. 
        Functions $\phi\in \mathcal{H}_N$ that are simultaneous
        eigenfunctions of all the Hecke operators are called Hecke
        eigenfunctions, and a basis consisting of Hecke
        eigenfunctions is called a Hecke basis.
        By definition $A\in C(N,A)$, consequently any Hecke
        eigenfunction is in particular also an eigenfunction of
        $U_N(A)$.

\section{Reduction to lower dimension}
    Let $A\in\Sp(4,\bbZ),\;A=I\pmod{2}$, with 4 distinct
    eigenvalues.
    Further assume that the vector space $\Q^4$,
    decomposes into two (rational) symplectic subspaces, invariant under the
    action of $A$. In this case, the Hecke group $C(N,A)=\bar{C}_1(N,A)\times\bar{C}_2(N,A)$ is a
    direct product, each term $\bar{C}_i(N,A)$, can be identified
    with a lower dimensional Hecke group of a corresponding matrix in $\SL(2,\bbZ/N\bbZ)$.
    Correspondingly, the Hecke operators and eigenfunctions are a tensor product of the appropriate
    lower dimensional Hecke operators and eigenfunctions.
\subsection{Reduction of Hecke group}
    For each invariant subspace, take a symplectic basis
    $e_i,f_i\in\Q^{4}$,
    that is $\omega(e_i,e_j)=\omega(f_i,f_j)=0$ and $\omega(e_i,f_j)=\delta_{i,j}$.
    For a sufficiently large prime $N$, take $\bar{e_i},\;\bar{f_i}\in
    (\bbZ/N\bbZ)^4$, through reduction of $e_i$ and $f_i$ modulo $N$ respectively. This induces a
    decomposition of $(\bbZ/N\bbZ)^4=E_1\oplus E_2$ into two orthogonal
    symplectic subspaces invariant under the action of $A\!\!\mod{N}$.

    For $\vec{n}\in\bbZ^4$, denote by $(\bar{n}_1,\bar{n}_2)\in(\bbZ/N\bbZ)^2\times(\bbZ/N\bbZ)^2$ the
    restriction of $\vec{n}\mod{N}$, to $E_1\oplus E_2$ in the
    symplectic basis. Since this is a symplectic decomposition then
    for any $\vec{n},\vec{m}\in\bbZ^{4}$
    \begin{equation}\label{eSYM}
    \omega(\vec{n},\vec{m})=\omega(\bar{n}_1,\bar{m}_1)+\omega(\bar{n}_2,\bar{n}_2)\pmod{N}.
    \end{equation}

    This decomposition induces a map
    \[i_N:\SL(2,\bbZ/N\bbZ)\times \SL(2,\bbZ/N\bbZ)\hookrightarrow \Sp(4,\bbZ/N\bbZ).\]
    If we denote by $A_i\in\SL(2,\bbZ/N\bbZ)$ the restriction of
    $A\pmod{N}$ to $E_i$ in the symplectic basis, then
    $i_N(A_1,A_2)\equiv A\pmod{N}$.
    \begin{lem}
        Let $\bar{C}_i(N,A)\subseteq \SL(2,\bbZ/N\bbZ)$ be
        maximal commutative subgroups which include $A_i$, $i=1,2$ respectively. Then
        the map $i_N$ defined above, maps the group
        $\bar{C}_1(N,A) \times \bar{C}_2(N,A)$
        isomorphically onto $C(N,A)$.
    \end{lem}
    \begin{proof}
        The map
        $i_N:\bar{C}_1(N,A)\times\bar{C}_2(N,A)\hookrightarrow C(N,A)$, is clearly injective.
        Thus, it is sufficient to show that this map is onto.
        Let $B\in\Sp(4,\bbZ/N\bbZ)$, be a matrix that commutes with
        $A\pmod{N}$. We can assume $N$ is large enough so that $A\pmod{N}$
        has 4 distinct eigenvalues in $F_{\!\!N^2}$.
        Consequently, the spaces $E_i$ are also invariant under
        the action of $B$. Thus if we denote by
        $B_i\in\SL(2,\bbZ/N\bbZ)$ the restriction of $B$ to $E_i$,
        in the symplectic basis, then $i_N(B_1,B_2)=B$.
    \end{proof}
    \subsection{Quantization of Hecke group}
    Let $T_N^{(1)}(\cdot), U_N^{(1)}(\cdot)$ be the
    quantized elementary observables and propagators for
    $\bbT^{2}$.
    For $\vec{n}\in\bbZ^{4}$, identify $\bar{n}_1,\bar{n}_2$
    defined above with elements of $(\bbZ/2N\bbZ)^2$ by requiring
    \begin{equation}\label{eSYM:2}
    \bar{n}_1=(n_1,n_3),\; \bar{n}_2=(n_2,n_4)\pmod{2}.
    \end{equation}
    We can also identify $B_i\in\bar{C}_i(N,A)$, as
    elements of $\SL(2,\bbZ/2N\bbZ)$ by requiring them to be congruent to
    $I$ modulo $2$.
    \begin{prop}\label{pTENS}
        There is a unitary mapping \[\calU:L^2(\bbZ/N\bbZ)^2\rightarrow
        L^2(\bbZ/N\bbZ)\otimes L^2(\bbZ/N\bbZ),\]
        such that:
        \begin{enumerate}
        \item For any
        $\vec{n}\in\bbZ^4$,
        \[\calU T_N(\vec{n})\calU^{-1}=T^{(1)}_N(\bar{n}_1)\otimes T^{(1)}_N(\bar{n}_2).\]
        \item For any $B\in C(N,A)$,
        \[\calU U_N(B)\calU^{-1}=U^{(1)}_N(B_1)\otimes U^{(1)}_N(B_2).\]
        \end{enumerate}
    \end{prop}
    \begin{proof}
    It is easily verified from \ref{eSYM}, \ref{eSYM:2} that $\widetilde{T}_N(\vec{n})=T^{(1)}_N(\bar{n}_1)\otimes
    T^{(1)}_N(\bar{n}_2)$ obey the same relation as
    in proposition \ref{pTN:1}.
    Thus, from uniqueness they are unitarily equivalent.

    As for the second part, recall $U^{(1)}_N(B_1)$ and $U^{(1)}_N(B_2)$ both satisfy the Egorov
    identity, and we showed that $\calU T_N(\vec{n})\calU^{-1}=T^{(1)}_N(\bar{n}_1)\otimes
    T^{(1)}_N(\bar{n}_2)$. Consequently, if we define
    $\tilde{U}_N(B)=\calU^{-1}U^{(1)}_N(B_1)\otimes U^{(1)}_N(B_2)\calU$,
    then $\tilde{U}_N(B)$
    satisfies the Egorov identity as well:
    \[\tilde{U}_N(B)^{-1}T_N(\vec{n})\tilde{U}_N(B)=T_N(\vec{n}B).\]
    Thus, from the uniqueness of the quantization
    \[\calU U_N(B)\calU^{-1}=U^{(1)}_N(B_1)\otimes U^{(1)}_N(B_2).\]
    \end{proof}
    \begin{rem}
        In fact, one can use this procedure to define the phases
        for the quantization of the Hecke group. Consequently, if
        the phases for the quantized two dimensional
        maps are chosen so that the quantization is
        multiplicative (as done in \cite{KR2}), then the quantization of the Hecke group
        would be multiplicative as well.
    \end{rem}

    \subsection{Hecke eigenfunctions}
    Let $\set{\phi^1_j}_{j=1}^N,\set{\phi^2_j}_{j=1}^N\in L^2(\bbZ/N\bbZ)$, be
    joint eigenfunctions of all the Hecke operators
    $U_N^{(1)}(B_i),\;B_i\in C_i(N,A)$ $i=1,2$
    respectively. Correspondingly,
    $\phi_{j_1,j_2}=\calU^{-1} (\phi_{j_1}^1 \otimes \phi_{j_2}^2)$,
    are eigenfunctions of all the operators
    $\set{U_N(B)|B\in C(N,A)}$, and hence $\set{\phi_{j_1,j_2}}_{j_1,j_2=1}^N$
    is a Hecke basis of $L^2(\bbZ/N\bbZ)^2$. In this basis, the matrix elements of the elementary
    observables $T_N(\vec{n})$, are given by
    \begin{equation}\label{eRED:1}
    \langle T_N(\vec{n})\phi_{j_1,j_2},\phi_{j_1,j_2}\rangle=
    \langle T^{(1)}_N(\bar{n}_1)\phi^1_{j_1},\phi^1_{j_1}\rangle
    \langle T^{(1)}_N(\bar{n}_2)\phi^2_{j_2},\phi^2_{j_2}\rangle.
    \end{equation}
    \begin{rem}
        The joint eigenspaces 
        of all the operators
        $U_N^{(1)}(B_i),\:B_i\in C_i(N,A)$, are one dimensional (except for the eigenspace
        corresponding to the trivial character) \cite{K}. Correspondingly,
        any Hecke eigenfunction is of the form defined above,
        except for Hecke eigenfunctions corresponding to the
        trivial character, which are of the form
        $\phi=\calU^{-1}(a\phi_1^1\otimes\phi_1^2+b\phi_2^1\otimes\phi_2^2)$,
        where $\phi_1^i,\phi_2^i\in L^2(\bbZ/N\bbZ)$ correspond to the trivial
        character, and $|a|^2+|b|^2=1$.
    \end{rem}

    \section{Variance of matrix elements}
    We now turn to prove theorem \ref{t:1}. We will concentrate on
    the elementary observables $f(\vec{x})=e^{2\pi\vec{n}\cdot
    \vec{x}}$ with corresponding quantum operators
    $\Op_N(f)=T_N(\vec{n})$, and calculate their variance in the Hecke basis.

    Following the construction of the Hecke eigenfunctions and
    matrix elements described in the previous section, it is
    sufficient to understand the distribution of the matrix
    elements for the cat map on $\bbT^2$.
    In the following proposition, we summarize some results
    regarding the matrix elements of elementary observables in the Hecke basis on the
    2-torus.
    \begin{prop}\label{pQV:1}
    Let $A\in \SL(2,\bbZ/N\bbZ)$ with 2 distinct eigenvalues.
    Let $\bar{C}(N,A)\subseteq \SL(2,\bbZ/N\bbZ)$ be
    the corresponding Hecke group, and
    $\set{\phi_i}_{i=1}^N\in L^2(\bbZ/N\bbZ)$ the corresponding Hecke
    eigenfunctions. For $\bar{n}\in (\bbZ/2N\bbZ)^2$ that is not
    an eigenvector of $A\pmod{N}$:
    \begin{enumerate}
    \item The second moment of the corresponding operator is
    \[\frac{1}{N}\sum_{j=1}^N |\langle T^{(1)}_N(\bar{n})\phi_j,\phi_j\rangle|^2=\left\lbrace\begin{array}{cl}
     1 & \bar{n}=0\pmod{N}  \\
      \frac{1}{N}+O(\frac{1}{N^2}) & \bar{n}\neq 0\!\!\!\pmod{N} \\
    \end{array}\right.\]
    \item
    For $\phi,\phi'$ Hecke eigenfunctions
    corresponding to the trivial character:
    \[|\langle
    T^{(1)}_N(\bar{n})\phi,\phi'\rangle|=\left\lbrace\begin{array}{cc}
     1 & \tilde{n}=0\pmod{N} \\
     O(N^{-1/2}) & \bar{n}\neq 0 \pmod{N}\
    \end{array}\right.\]
    \end{enumerate}
    \end{prop}
    For proof we refer to \cite{KR1}.
    We now give the proof of theorem \ref{t:1}:
    \begin{proof}
        Let $\set{\phi_{j_1,j_2}}_{j_1,j_2=1}^{N}$ be the Hecke basis
        constructed in the previous section.
        From the formula given for the matrix elements in
        \ref{eRED:1}, we can rewrite the quantum variance as
        \[S_2(f,N)=\frac{1}{N}\sum_{j_1=1}^{N}|\langle
        T^{(1)}_N(\bar{n}_1)\phi^1_{j_1},\phi^1_{j_1}\rangle|^2\frac{1}{N}\sum_{j_2=1}^{N}|\langle
        T^{(1)}_N(\bar{n}_2)\phi^2_{j_2},\phi^2_{j_2}\rangle|^2.\]
        Now if $\vec{n}\in Z_1$, then it is a linear combination
        of $e_1,f_1$, and hence for $N$ sufficiently large $\bar{n}_1\neq 0,\;\bar{n}_2=0\mod{N}$.
        In the same way if $\vec{n}\in Z_2$ then $\bar{n}_1=0,\;\bar{n}_2\neq
        0\mod{N}$, and if $\vec{n}\not\in Z_1\cup Z_2$ then $\bar{n}_1,\bar{n}_2\neq 0\mod{N}$.
        Next, since $A$ has no rational eigenvectors we can assume $N$ is sufficiently large
        so that $\bar{n}_1,\bar{n}_2$ are not eigenvectors of $A_1,A_2$ respectively.
        The proof now follows directly from the first part of proposition \ref{pQV:1}.
    \end{proof}
    \begin{rem}
        Note that if we take a different Hecke basis from the one
        constructed in the previous section, then we only change
        the elements in the sum corresponding to the trivial
        character. These elements, from the second part of proposition \ref{pQV:1},
        contribute $O(\frac{1}{N^2})$ (respectively
        $O(\frac{1}{N^3})$ if $\vec{n}\not\in Z_1\cup Z_2$). Therefor this result holds for any
        Hecke basis.
    \end{rem}

    \section{Discussion}
    \subsection{Product behavior}
    The main ingredient in the proof of theorem \ref{t:1}, is the
    fact that the quantized system can be identified as a tensor
    product of two quantized two dimensional systems, and that there are observables
    whose quantization is trivial on one of the factors.

    This behavior is
    clearly to be expected if the classical four dimensional
    system is a product of two different two dimensional systems
    (e.g., considering the
    four dimensional torus as a product $\bbT^4=\bbT^2\times\bbT^2$, and
    constructing an element $A\in Sp(4,\bbZ)$ by taking two elements
    $A_1,A_2\in Sp(2,\bbZ)$ acting on each factor).
    In such a system, when looking at observables that are constant on one factor,
    the fluctuations of their quantization
    only come from the nonconstant factor, and hence the anomalous rate of decay.

    However, it is interesting to note that this behavior can also
    occur for systems that do not factor in terms of their
    classical dynamics. The condition for the classical dynamics to
    factor, is that there is a symplectic rational invariant
    decomposition such that the lattice $Z_1\oplus Z_2=\bbZ^4$ (where $Z_1,Z_2$ are as in theorem
    \ref{t:1}). In some cases, even though the space $\Q^4$
    decomposes into two invariant rational symplectic subspaces, the lattice $Z_1\oplus Z_2\subset \bbZ^4$
    is a proper sublattice of finite index and the classical dynamics does not factor.
    Nevertheless, for any prime $N$ the vector space $(\bbZ/N\bbZ)^4$ decomposes
    in to two invariant symplectic subspaces, and the quantized system would still behave like that of a product of two systems.

    \begin{rem}
    It is worth mentioning that even if the matrix $A$ does not factor
    over the rationals, it does factor over a quadratic extension,
    and hence for half the primes it would factor
    modulo $N$. Consequently, for these primes we would still have
    product behavior, that is the quantized system is a
    tensor product of two quantized two dimensional systems.
    However, in this case there can be no classical observables
    whose quantization is trivial on one of the factors (for infinitely many primes).
    \end{rem}

    \subsection{Limit distribution}
    The numerical evidence presented in figures \ref{f:1},\ref{f:2} suggests that for the degenerate case (where the rate
    of decay is $\frac{1}{N}$) the limit distribution follows the
    semicircle law, while otherwise it is given by the product of
    two random variables with semicircle distribution.

    In fact, assuming the validity of the Kurlberg-Rudnick conjecture
    regarding the limit distribution of matrix elements on the two dimensional
    torus \cite{KR2}, these results can be derived directly from the product behavior of the quantized
    system (\ref{eRED:1}).

    \subsection{Rate of decay}
\begin{figure}[t]
        \includegraphics[width=4in ]{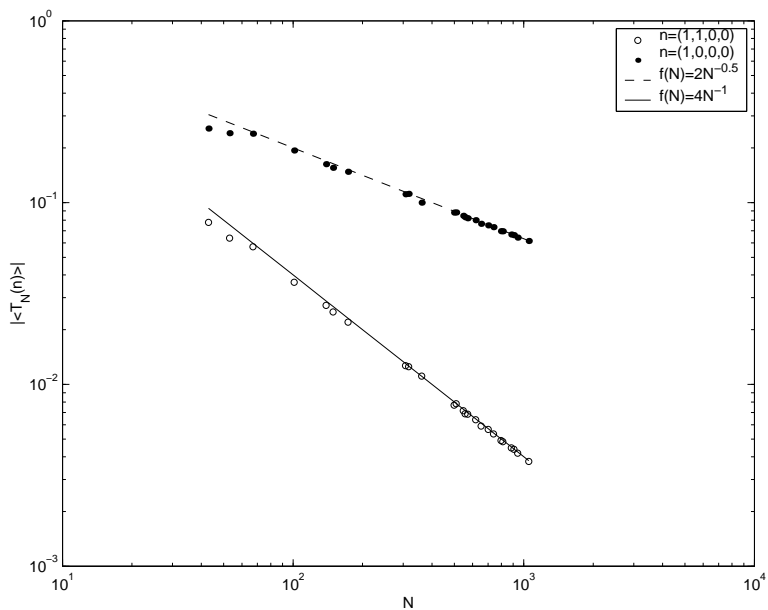}
        \caption{  The values of $\max_i |\langle T_N(\vec{n})\phi_i,\phi_i \rangle|$
        for $\vec{n}=(1,0,0,0),(1,1,0,0)$ and values of N between 100 and
        1000. Plotted together with the predicted decay of $ 2N^{-0.5},4N^{-1}$ respectively, on a log-log scale. }\label{f:3}
    \end{figure}

        The numerical evidence presented in figure \ref{f:3},
        imply a bound on the individual matrix elements:
        \begin{equation}\label{eBND}|\langle T_N(\vec{n})\psi_i,\psi_i\rangle| \leq
        \left\lbrace\begin{array}{cc}
        \frac{2}{N} & \vec{n}\in Z_1\cup Z_2\\
        \frac{4}{N^2} & \mbox{ otherwise}\
        \end{array}\right.
        \end{equation}
        Given the Kurlberg-Rudnick rate conjecture (which is now a theorem \cite{GH2})
        the above bound (\ref{eBND}), can be derived from
        (\ref{eRED:1}) as well.

\clearpage



\providecommand{\bysame}{\leavevmode\hbox
to3em{\hrulefill}\thinspace}
\providecommand{\MR}{\relax\ifhmode\unskip\space\fi MR }
\providecommand{\MRhref}[2]{%
  \href{http://www.ams.org/mathscinet-getitem?mr=#1}{#2}
} \providecommand{\href}[2]{#2}

\end{document}